\documentclass[usenatbib]{mnras}
\usepackage[T1]{fontenc}
\usepackage{amsmath}
\usepackage{multirow}
\usepackage{graphicx}
\providecommand{\tabularnewline}{\\}
\begin{document}
\title{Solar and stellar activity cycles — no synchronization with exoplanets}
\author[V.~Obridko et al.]{V.N.~Obridko$^{1,4}$\thanks{obridko@izmiran.ru}, M.M.~Katsova$^{2}$, D.D.~Sokoloff$^{1,3,5}$ \\
$^{1}$IZMIRAN, 4, Kaluzhskoe  Shosse, Troitsk, Moscow, 142190, Russia\\
$^{2}$Sternberg State Astronomical Institute, Lomonosov Moscow State University, 13, Universitetsky prosp., Moscow, 119991, Russia\\  
$^{3}$Department of Physics, Lomonosov Moscow State University, Moscow, 119991, Russia\\
$^{4}$Central Astronomical Observatory of the Russian Academy of Sciences at Pulkovo, St.Petersburg, Russia\\
$^5$Moscow Center of Fundamental and Applied Mathematics, Moscow, 119991, Russia
}

\date{
Accepted April 2022}
\maketitle

\begin{abstract}
Cyclic activity on the Sun and stars is primarily explained by generation of the magnetic field by a dynamo mechanism, which converts the energy of the poloidal field into the energy of the toroidal component due to differential rotation. There is, however, an alternative point of view, which explains the field generation by gravitational influence of the planetary system and, first of all, Jupiter. This hypothesis can be verified by comparing the characteristics of exoplanets with the activity variations on their associated stars. We have performed such a comparison and have drawn a negative conclusion. No relationship between the gravitational influence of the exoplanets and cycle of the host star could be found in any of the cases considered. Moreover, there are reasons to believe that a strong gravitational influence may completely eliminate cyclic variation in  stellar activity.
\end{abstract}


\section{Introduction}
The best-known phenomenon of solar activity is the 11-year cycle. Its origin is thought to be associated with self--excitation of the solar magnetic field somewhere in the solar interior due to the electromagnetic induction effect known as the solar dynamo {\bf \citep[e.g.][]{Cetal18, Ketal22}}. This idea is widely accepted in the expert community, although the particular features of the process still remain a matter of scientific discussion. An interesting point, however, is that the length of a solar cycle (about 11 years) is remarkably close to the orbital period of Jupiter. At first glance, it is very tempting to see a connection between the above two quantities and suggest that the orbital period of Jupiter somehow determines the very existence of the solar activity cycle or at least its length. This idea has been offered many times in different circumstances and in different forms since the 19th century. To save space, we will not give here an extended historical overview of the idea, confining ourselves to a few more or less arbitrarily chosen references \citep{Cetal12, Betal05, Aetal12}. For more references, see \cite{Stetal2019}. This paper, as well as \cite{Stetal2018,Stetal2020, Stetal2021}, presents the current state of the problem. An important addition to the initial idea here is that the discussion involves the influence of the other planets of the solar system on periodic variations of the solar activity.

Many experts involved in the study of the solar activity do not accept enthusiastically the idea of planetary synchronization of the solar cycle, strongly preferring the dynamo explanation {\bf \citep[e.g.][]{Hetal19, Retal22}.} The point is that Jupiter is quite remote from the Sun, and its influence on the flows in the solar interior is much weaker than various convective effects therein. It is, however, quite difficult to prove that a long-term action of a weak force cannot somehow affect the flows in the solar interior and, thus, participate in the formation of the solar cycle. On the other hand, the statement that the similarity of the length of the solar cycle and the orbital period of Jupiter are a mere coincidence seems to require some justification.

In our opinion, the justification required can be obtained by showing that other stars more or less similar to the Sun do not demonstrate such a similarity between their activity cycles and the orbital rotation of exoplanets or even the rotation period of a stellar companion. Until recently, we did not have sufficiently reliable data on the stellar activity and the orbital periods of exoplanets to make such a comparison. At present, the level of data accumulated in the respective areas of research allows us to perform this comparison. This is the purpose of our article. Realizing that various effects of the observational selection complicate the comparison, we will demonstrate that the observational data accumulated are instructive enough for at least preliminary conclusions. E.g, only four of the 111 program stars of the HK Project \citep{Baletal1995} host exoplanets and demonstrate  a cyclic activity: HD 26965, period of cycle  $10.1\pm0.1$ yrs --  a Neptune-like planet with $P_{\rm orb} = 42.3$ days; HD 3651, period of cycle $13.8\pm0.4$ yrs  -- a gas giant with $P_{\rm orb} = 62.3$ days; HD 190007, period of cycle  13.7 yrs/5.3 yrs  -- a Neptune-like planet with $P_{\rm orb}$ = 11.7 days; and HD 206860, period of cycle 6.2 yrs -- a gas giant with $P_{\rm orb}$ = 20692 yrs (!) 

Thus, in none of these cases is there any close coincidence of the cycle period with the orbital period of the satellite planet. Note that the planetary hypothesis attempts to explain the 11-year cycle and not the experimentally found 22-year cycle with the polarity changing every 11 years\footnote{We are fortunate that stellar activity tracers are  not sensitive to the magnetic polarity.}. This is due to the fact that the hypothesis is based on the concept of a gravitational effect of Jupiter  
on the solar plasma.  Let us consider this in more detail in relation to the solar planetary system.

\section{Gravitational potential from an individual planet on the surface of a star}

Assuming that the orbital plane of the planet is close to the equatorial plane of the star, the gravitational potential on the star surface is
\begin{equation}
V=-\frac{\gamma Mr^2}{2R^3}(3\cos^2\phi-1).
\end{equation}
\noindent Here, $\gamma=6.67430(15)\cdot10^{-11}$ m$^3$ s$^{-2}$ kg$^{-1}$  or N m$^2$ kg$^{-2}$, M is the mass of the planet, r is the radius of the star, R is the distance from the star to the planet, and $\phi$ is the latitude.

Let us estimate the gravitational potential from Jupiter. 
The mass of Jupiter is $1.898 \cdot 10^{27}$ kg, the radius of the Sun is $7\cdot10^8$ m, and the orbit radius of Jupiter is $R=7.78 \cdot 10^{11}$ m.

The gravitational potential created by Jupiter on the surface of the Sun is 
\begin{equation}
V=-6.5\cdot10^{-2}(3\cos^2\phi-1)\ \ N/kg,
\end{equation}
the radial force is
\begin{equation}
F_r=\frac{\partial V}{\partial r}=\frac{4}{r}V=-4\cdot10^{-10}(3\cos^2\phi-1)\ \ Nm^{-1}/kg
\end{equation}
and the meridional force is
\begin{equation}
F_{\phi}=\frac{1}{r}\frac{\partial V}{r\partial\phi}=-6\cdot10^{-10}\sin2\phi\ \ Nm^{-1}/kg.
\end{equation}

With the density values at the top of the convection zone, the acceleration is $\approx 10^{-8}$ m/s$^2$. At such acceleration and at the characteristic times of processes in the convection zone equal to several days, the radial component of the velocity is insignificant. However, in the meridional flow, the characteristic times are several years (up to half a cycle); therefore, the variation in the velocity of the meridional flow, in principle, may reach several m/s for 2-3 years. This is comparable with the measured values of 10-15 m/s and may slightly change the height of the cycle \citep{Getal2009, Obretal2012, Cetal12,  Ge2013}. However, it should be noted that the obtained values are insufficient for the occurrence of the 11-year cycle. 
Here we have to stress that we discuss here the planetary influence as the driver which determines the very existence and duration of solar cycle. Of course, solar and stellar dynamo is a complicated nonlinear process and it is more than possible that even weak parametric variations can results in various modifications of dynamo solutions \citep[e.g.][]{mps02}. To be specific, we do not consider such options systematically in this paper.
For the solar activity cycle, one can talk about an additional modulation of the height of the 11-year cycle due to the proximity (perhaps casual) of the orbital period of Jupiter and the 11--year cycle generated by the dynamo processes. In principle, planetary effects seem to be sufficient strength to affect meridional circulation and associated properties of solar cycle. Further modelling of such options looks attractive however it is  beyond of the aims of this paper.

In any event, two conditions must be met. Firstly, the gravitational influence cannot be too strong, otherwise it will come into conflict with the dynamo process and the cyclic activity will be disturbed,  e.g. the freezing of differential rotation to solid body rotation, turning off the $\Omega$ effect, and transforming  conventional $\alpha \Omega$-dynamo on $\alpha^2$-one can be considered. The same disruption can occur if the orbital period of a massive nearby planet differs essentially from the natural cycle of the magnetic-field generating dynamo.

Next, we investigate how much the gravitational potential on stars with exoplanets differs from that in the solar system (if a given star has several exoplanets, we choose the planet that ensures the highest potential).

\section{Observational basis for stellar cycles}

Stellar cyclic activity has been revealed in the course of long-term monitoring of 
chromospheric 
variations in main-sequence stars \citep{Baletal1995}. Since the broad  spectral lines of Ca II H (3968 A) and K (3934 A) are clearly visible in the solar-like stars and show emission peaks in the line cores, and this emission is largely magnetically heated, the fluxes in these spectral lines serve as an important tracer of the stellar activity. Uniformly calibrated long-term records of this proxy were obtained within the framework of the long-standing HK Project for 111 stars of the spectral types F2--M2  on or near the main sequence. \cite{Baletal1995} revealed a variation pattern in the rotation and chromospheric activity of G0--K5 stars on an evolutionary timescale, in which high levels of activity with rare cyclic variations were recorded in young fast-rotating stars; moderate activity and random smooth cycles were revealed in stars of intermediate age; and slowly rotating stars of solar age and older demonstrated the lowest levels of activity during smooth cycles with occasional epochs of Maunder-like minima. Some of the oldest stars may have ceased cyclic dynamo activity \cite{MS17}. It was noted that certain periodic variations are similar to the solar cycle. They were classified as “Excellent” and “Good” and were observed in 21 stars, including the Sun. Not very clearly defined periodicity was classified as “Fair” and “Poor”; it was recorded in 25 stars. The rest of the stars demonstrated different degrees of variability classified as follows: “Var” means significant variability on the timescales longer than 1 yr but much shorter than 25 yr without pronounced periodicity. “Long” means significant variability on the timescales longer than 25 yr. Note that some records show secular change over 25 yr which suggests that the cycle period, if any, is longer than 50 yr. 
In some HK Project stars, the so-called “Flat” activity was identified, when the 
index of chromospheric activity remained constant with time. \cite{Betal22} re-classified the extended record of time variation of the stellar activity and provided a refined classification of the behavior of activity, as well as more precise periods of cycles for some HK Project stars. Taking into account that the comparison under discussion is quite new, we try to present here the idea of research only and  choose largely a single source of cycle measurements \cite{Baletal1995}, with
supplements from \cite{Betal22}  and using other sources sparingly.  We fully appreciate that the idea presented here deserves further development which have to include in particular \cite{Oetal09, Letal11, Oetal16,  Letal16}.

Data on exoplanets associated with the HK Project stars were taken from the databases of Extrasolar planet catalogues:
https://exoplanets.nasa.gov/discovery/exoplanet-catalog/  and
http://exoplanet.eu/catalog/

\section{Results}

We found out that 16 of the 111 program (known as HK Project) stars \citep{Baletal1995} with different characteristics of the cyclic chromospheric activity have planets of the {\bf mass} of Jupiter (gas giant) or {\bf super-Earths}, as well as two planetary systems, see Table 1. (The rotation periods of the host stars are given in brackets). Combining the results, we found the following sets of stars with identified activity: Excellent ($P_{\rm rot} = 43$ d) -- 1; Good ($P_{\rm rot} = 44$ d) -- 1; Poor (5 d) -- 1; Fair/Poor (29 d) -- 1; Var (12 d) -- 4; Long (26 d, 17 d) -- 2; Flat (34 d, 38 d, 9 d) -- 3. Rotation periods here are taken from \cite{Betal96} and deserve further confirmation. Fortunately, we need them primary to give a hint of stellar age. Another motivation relevant for the topic is that if $P_{\rm rot}$ is close to the orbital period $P_{\rm orb}$ (say, HD 26965) one may expect that the star is tidally locked with the planet and the cycle properties may be tidally forced. This gravitationally based effect is relevant for the field under discussion however is beyond the scope of our paper.

A situation more or less similar to that observed on the Sun is revealed only on three stars:  HD 22049, HD 190360, and HD 190406. The star HD 190360 has no cycle, the star HD 190406 has a cycle lasting 2 years, i.e., 20 times less than its rotation period, and the cycle on the star HD 22049 is not well-defined (unsettled). On HD 126053 and HD 206860, the potential is so weak that the planets cannot interfere with the dynamo. HD 176051AB is actually a binary system. \cite{Metal10} do not known around which component the planet b, detected by astrometry, is orbiting.
The two components are HD 176051A (1.07 Solar mass F9 V star, V = 5.28)  and
HD 176051B (0.71 Solar mass K1 V star, V = 7.82).
Exoplanet.eu takes the (rather artificially) mean value $(M_A + M_B)/2$ for the stellar mass.
If the planet is orbiting the 1.07 Solar mass A component, the planet mass is 2.26 Mj and $R_{\rm orb} = 2.02$ AU.

On the rest of the stars under examination, the planetary effect is so strong that one may expect that
disrupted the entire structure of the differential and meridional flows, so that a normal operation of the dynamo is out of the question. The latest data from the ongoing HK Project \cite{Betal22} support the above results (see Table 2). 

The general conclusion is as follows: in most cases, the planets either do not have any effect at all, or disrupt the regular structure of the dynamo. The impact is possible in rare cases when the rotation period of a massive planet is close to the period of the cycle generated by the dynamo independently of the planets. In this case, the planet may produce a modulating effect by influencing the heights of successive cycles.

\begin{table*}
\caption{Stars with planets and the type of activity known from HK Project.  $P_{\rm rot}$ is the rotation period, $P_{\rm cyc}$ is the cycle length, $M_p$ is the mass of the planet (stands in the rows for planet), $V_p$ (given in bold) is the ratio of the planetary influence on the star to the influence of Jupiter on the Sun (stands in the stellar row), $R_p$ is the radius  of the planet, $P_{\rm orb}$ is the orbital period, and $R_{\rm orb}$ is the orbital radius in astronomical units (AU). The activity according to HK data is denoted as follows: E stands for Excellent, G --- Good, F --- Flat, L --- Long, and V --- Var. $\ne$ means that $P_{\rm orb} \ne P_{\rm cyc}$, Me is the mass of the Earth, Re is the radius of the Earth, Mj is the mass of Jupiter, Rj is the radius of Jupiter, 
b means a binary system, n marks the stars where the planetary hypothesis suggests the existence of an activity cycle, while observations do not reveal any cycle, and * means that the orbital period is estimated from the Kepler's law.}
\begin{tabular}{|l|l|l|l|l|l|l|l|l|}
\hline
Name & $P_{\rm rot}$ & $P_{\rm cyc}$ &$V_p$  ($M_p$) & $R_p$ & $P_{\rm orb}$ & $R_{\rm orb}$ &  \cr
\hline  
HD 3651 = 54 Psc (K0 V)& 44 d & 13.8 yrs & \Large{\boldmath$10^3$} &&&& G &\cr
HD 3651 b && & 0.228 Mj & 0.899 Rj & 62.3 d & 0.295 AU & $ \ne $ \cr 
HD 3651 B &&& $53\pm 15$ Mj & 0.8 Rj && 476 AU &\cr
\hline
HD 10700 = $\tau$ Cet (G8 Vp) & 34 d &   & \Large{\boldmath$2\times 10^2$}&&&& F \cr
$\tau$ Cet e &&& 3.93 Me & 1.18 Re  & 162.9 d & 0.538 AU &  \cr
$\tau$ Cet f &&& 3.93 Me & 1.18 Re & 1.7 yrs & 1.334 AU &  \cr
$\tau$ Cet g &&& 1.75 Me & 1.81 Re & 20 d & 0.133 AU & n \cr
$\tau$ Cet h &&& 1.83 Me & 1.19 Re & 49.4 d & 0.243 AU & \cr
\hline
HD 22049=$\epsilon$ Eri (K2 V) & 12 d &  & \Large{\boldmath$\approx 1$} &&&& L \cr
$\epsilon$ Eri b &&& 0.78 Mj & 1.24 Rj & 7.4  yrs & 3.5 AU & n \cr
\hline
HD 26965 =$o^2$ Eri (K1 V) & 43 d & 10.1 yrs &\Large{\boldmath$2\times 10^2$}  &&&& E \cr
HD 26965 b &&& 8.47 Me & 0.254 Rj &  42.4 d & 0.215 AU &  $\ne $ \cr
\hline
HD 89744 (F6-7 V) & 9 d &  & \Large{\boldmath$9 \times 10^3$} &&& & F \cr
HD 89744 b &&& 8.35 Mj & 1.12 Rj & 256.8 d & 0.917 AU & \cr
HD 89744 c &&& $5.36 \pm 4.57$ Mj &  & 6974 d & 8.3 AU & n \cr
\hline
HD 95735=GJ 411 (M2.1 Ve) & 53 d & & \Large{\boldmath$4 \times 10^2$} &&&& V \cr
Lalande 21185 b=GJ 411 b &&& 2.69 Me & 1.45 Re & 12.9 d & 0.079 AU &  \cr
Lalande 21185 c=HD 95735 c &&& 18.05265 Me & 0.396 Rj & 8.7 yrs & 3.1 AU & n \cr
\hline
HD 115617=61 Vir (G6 V) & 29 d & & \Large{\boldmath$2 \times 10^4$}&&&& V \cr
61 Vir b &&& 5.1 Me & 2.11 Re & 4.2 d & 0.050 AU &   \cr
61  Vir c &&& 18.2 Me & 0.398 Rj & 38 d & 0.217 AU & n \cr
61 Vir d &&& 22.9 Me & 0.456 Rj & 123 d & 0.476 AU & \cr
\hline
HD 126053 (G3 V) & 22 d & 22 (?) yrs & \Large{\boldmath$10^{-7}$} &&&& $ \ne $ \cr
HD 126053 B &&& $35 \pm 15$ Mj & 0.9 Rj & $ 10^6$ yrs*  & 2630 AU &  \cr
\hline
HD 141004=GJ 598=$\lambda$ Ser (G0 V) & 26 d & &\Large{\boldmath$3 \times 10^3$} &&&& L \cr
HD 141004 b &&& 13.65 Me & 0.366 Rj & 15.5 d & 0.124 AU & n \cr
\hline
HD 143761 = $\rho$ CrB (G2 V)  & 17 d & & \Large{\boldmath$3 \times 10^4$} & &&& L  \cr
$\rho$ CrB b &&& 1.0449 Mj & 1.23 Rj & 39.8 d & 0.220 AU & \cr
$\rho$ CrB c &&& 25 Me & 0.48 Rj & 102.5 d & 0.412 AU & n  \cr
\hline
HD 176051AB (G0 V) & 16 d & 10(?) yrs &&& & & b \cr
HD 176051 b &&& 1.5 Mj && $1016.0 \pm 40.0$ d  & 1.76 AU & $ \ne $ & \cr
\hline
HD 190007=GJ 775 (K4 V) & 29 d & 13.7 yrs (?) & \Large{\boldmath$9 \times 10^3$}&&&& F  \cr
HD 190007 b &&& 16.46 Me & 0.375 Rj & 11.7 d & 0.092 AU & $\ne $ &\cr
\hline
HD 190360 (G6 IV) & 38 d && \Large{\boldmath$4 \times 10^4$} &&&& F \cr
HD 190360 b &&& 1.54 Mj & 1.21 Rj & 8 yrs & 3.97 AU &  \cr
HD 190360 c &&& 19.069 Me & 0.409 Rj & 17.1 d &  0.134 AU & n \cr
\hline
1HD 190406=GJ 779=15 Sge (G1 V) & 14 d & 16.9 yrs &\Large{\bf 3} &&&& G \cr
HR 7672 b &&& 61.5 Mj & & 52 yrs* & 14 AU & $\ne $\cr
\hline
HD 206860=HN Peg (G0 V) & 5 d & 6.2  yrs & \Large{\boldmath$\approx 0$} &&&& $\ne$ \cr
HN Peg b & && 21.9987 Mj & 1.051 Rj & 20692.2 yrs &773 AU  & \cr
\hline
HD 217014=GJ 882=51 Peg (G5 V) & 37 d & & \Large{\boldmath$7 \times 10^5$}&&& & V  \cr
51 Peg b &&& 0.46 Mj & 1.27 Rj & 4.2 d & 0.053 AU & n \cr
\hline
\end{tabular}
\end{table*}

\begin{table*}
\caption{Stars with planets and the type of activity known from \protect\cite{Betal22}. 
The last row gives the data for the Sun as a star and Jupiter. Notations as in Table 1.
}
\begin{tabular}{|l|l|l|l|l|l|l|l|l|}
\hline
Name & $P_{\rm rot}$ & $P_{\rm cyc}$ & $V_p$ ($M_p$) & $R_p$ & $P_{\rm orb}$ & $R_{\rm orb}$ &  \cr
\hline  
HD 1461  (G3 V)& 29 d &  & \Large{\boldmath$10^4$} &&&& V &\cr
HD 1461 b && & 6.44 Me & 0.216 Rj & 5.8 d & 0.063 AU & n \cr 
HD 1461 c &&& 5.59 Me & 2.23 Rj & 13.5 d & 0.011 AU &\cr
\hline
HD 7924  (K0 V) & 35 d & 7.2 yrs  & \Large{\boldmath$2\times 10^3$} &&&& V  \cr
HD 7924 b &&& 6.357 Me & 0.214 Rj  & 5.4 d & 0.06 AU & $\ne$ \cr
\hline
HD 10697 (G3 Va) & 36 d &  & \Large{\boldmath$9 \times 10^1$} &&&& F \cr
HD 10697 b &&& 6.383 Mj & 1.13 Rj & 2.9  yrs & 2.14 AU & n \cr
\hline
HD 37124 (G4 IV-V) & 25 d &  & \Large{\boldmath$6\times 10^2$} &&& & V \cr
HD 37124 b &&& 0.675 Mj & 1.25 Rj & 154.4 d & 0.534 AU & \cr
HD 37124 c  &&& 0.652 Mj & 1.25 Rj & 2.4 yrs & 1.71 AU & n \cr
HD 37124 d &&& 0.696 Mj & 1.25 Rj & 5.1 yrs & 2.807 AU & \cr 
\hline
HD 178911B (M2.1 Ve) & 36 d & & \Large{\boldmath $3\times 10^4$} &&&& V \cr
HD 178911 B b  &&& 8.03 Mj & 1.12 Rj & 71.5 d d & 0.34 AU &  n \cr
\hline
HD 210277 (G8 V) & 41 d && \Large{\boldmath$10^2$} &&&& F \cr
HD 210277 b &&& 1.29 Mj & 1.22 Rj & 442.2 d & 1.13 AU &   n \cr
\hline
The Sun & 25 d & 11 yrs & \Large{\boldmath1} & && & E \cr
Jupiter & && 1 Mj &  1 Rj & 11.86 yrs &  5.204 AU &\cr 
\hline
\end{tabular}
\end{table*}

\section{Conclusion and Discussion}

Our conclusion is quite straightforward. We do not see in the data under discussion any support of the idea that the activity cycle in stars is the result of the planetary effect. We have to conclude that the coincidence between the orbital period of Jupiter and the solar activity cycle is purely accidental. At least we have to think so until a substantial number of planetary systems  (in addition to the case of Jupiter) are found, where the orbital period can be identified with the stellar activity cycle. Of course, we do not deny in principle that the orbital motion may contribute to the activity cycle of a star. It seems reasonable to believe that this may occur in close binaries. For example, \cite{mps02} investigated such effects to found out that it is quite difficult to affect the cycle substantially. 

We stress that our paper is a very initial, preliminary comparison of $P_{\rm cyc}$ and $P_{\rm orb}$ in context of possible gravitational effects in stellar dynamo. The number of systems  {\bf studied here} with both known $P_{\rm cyc}$ and $P_{\rm orb}$ is 9 only and 4 of them have less certain $P_{\rm cyc}$. Absence of observable cycles in another stars discussed is instructive in the very context of the paper however enlargement of the sample is very important to confirm our results and allow discussion of another gravitational effects. 

We mention in this context the case of $\epsilon$ Eri where the cycle was not seen in \cite{Baletal1995, Betal22}  however \cite{Jetal22} using long-term ZDI and HK data report two cycles 12.7 yr and 2.95 yr. The latter agrees with \cite{Cetal20} while the two are identical with cycles isolated by \cite{Metal13}. This result provides an additional confirmation for our conclusion because neither $P_{\rm cyc}$ matches $P_{\rm orb}$. Some additional data concerning cyclic stellar activity according to the HARPS planet search project may be obtained from \cite{Letal11} however their comparison with the data of HK project requires additional research  using in particular photometric sources
and other HK sources mentioned above. 

 In the framework of this  paper our aims are quite limited. We note however that the progress in exoplanet studies and stellar activity cycle observations opens a new areas for research, i.e. gravitational effects in dynamo. Here we can suggest a search  for exoplanetary systems with dynamo resonance effects, i.e. $P_{\rm cyc} = P_{\rm orb}$ or $P_{\rm orb} = 2 P_{\rm cyc}$. Another option is a search for exoplanetary systems where planetary forces involved are comparable with forces due to differential rotation \cite {Detal96, Baetal05, S11} or/and meridional flows. Both options are obviously out of the scope of this very paper.   

\section*{Acknowledgments}
VNO, MMK and DDS acknowledge the support of Ministry of Science and Higher Education of the Russian Federation under the grant 075-15-2020-780 (VNO and MMK) and 075-15-2022-284 (DDS). DDS thanks support by BASIS fund number 21-1-1-4-1. We acknowledge important comments from Dr Steven H. Saar.

Data availability statements.  Search for exoplanets around the HK-Project stars was carried out in databases of Extrasolar planet catalogues and NASA Exoplanet Archive
https://exoplanets.nasa.gov/discovery/exoplanet-catalog/  and
http://exoplanet.eu/catalog/. We used stellar activity data from \cite{Baletal1995, Betal22}. 
In this research we used of the SIMBAD database, operated at CDS, Strasbourg, France, and of NASA’s
Astrophysics Data System Bibliographic Services.

\bibliographystyle{mnras}
\bibliography{sample631}

\end{document}